\documentclass[12pt]{article}
\usepackage{graphicx}
\usepackage{bm}
\usepackage{amsmath}
\usepackage{times}
\usepackage[superscript,nomove]{cite}
\usepackage{setspace}
\usepackage{caption}
\usepackage{bibspacing}
\newenvironment{sciabstract}{%
\bf}
{}

\title{\large\vspace{-3.6cm}Topological charge-entropy scaling in kagome Chern magnet TbMn$_6$Sn$_6$\vspace{-0.2cm}}
\author
{\vspace{-0.1cm}\normalsize{Xitong Xu,$^{1,2}$ Jia-Xin Yin,$^{3}$ Wenlong Ma,$^{2}$ Hong-Ru Tian,$^{4,5}$ Xiao-Bin Qiang,$^{6}$ Huibin Zhou,$^{2}$}\\
\vspace{-0.15cm}\normalsize{Jie Shen,$^{1,7}$  Haizhou Lu,$^{6}$ Tay-Rong Chang,$^{4,5}$ Zhe Qu,$^{1,8\dag}$ Shuang Jia$^{2,9,10\ast}$}\\
\vspace{-0.15cm}\footnotesize{$^{1}$Anhui Key Laboratory of Condensed Matter Physics at Extreme Conditions,High Magnetic Field Laboratory, }\\
\vspace{-0.15cm}\footnotesize{HFIPS, Chinese Academy of Sciences, Hefei, Anhui 230031, China}\\
\vspace{-0.15cm}\footnotesize{$^{2}$International Center for Quantum Materials, School of Physics,}\\
\vspace{-0.15cm}\footnotesize{Peking University, Beijing 100871, China}\\
\vspace{-0.15cm}\footnotesize{$^{3}$Beijing National Laboratory for Condensed Matter Physics, Institute of Physics,}\\
\vspace{-0.15cm}\footnotesize{Chinese Academy of Sciences, Beijing 100190, China}\\
\vspace{-0.15cm}\footnotesize{$^{4}$Department of Physics, National Cheng Kung University, Tainan 701, Taiwan}\\
\vspace{-0.15cm}\footnotesize{$^{5}$Center for Quantum Frontiers of Research and Technology (QFort), Tainan 701, Taiwan}\\
\vspace{-0.15cm}\footnotesize{$^{6}$Department of Physics and Shenzhen Institute for Quantum Science and Engineering,}\\
\vspace{-0.15cm}\footnotesize{Southern University of Science and Technology, Shenzhen 518055, China}\\
\vspace{-0.15cm}\footnotesize{$^{7}$Science Island Branch of Graduate School,}\\
\vspace{-0.15cm}\footnotesize{University of Science and Technology of China, Hefei, Anhui 230026, China}\\
\vspace{-0.15cm}\footnotesize{$^{8}$CAS Key Laboratory of Photovoltaic and Energy Conservation Materials,}\\
\vspace{-0.15cm}\footnotesize{Hefei Institutes of Physical Science, Chinese Academy of Sciences, Hefei, Anhui 230031, China}\\
\vspace{-0.15cm}\footnotesize{$^{9}$Interdisciplinary Institute of Light-Element Quantum Materials and Research Center }\\
\vspace{-0.15cm}\footnotesize{for Light-Element Advanced Materials, Peking University, Beijing 100871, China}\\
\vspace{-0.15cm}\footnotesize{$^{10}$CAS Center for Excellence in Topological Quantum Computation,}\\
\vspace{-0.15cm}\footnotesize{University of Chinese Academy of Sciences, Beijing 100190, China}\\
\vspace{-0.15cm}\footnotesize{$^{\dag\ast}$E-mail:  zhequ@hmfl.ac.cn; gwljiashuang@pku.edu.cn}
}
\date{}

\begin{document}
\baselineskip21pt
\captionsetup[figure]{labelfont={bf},name={Fig.},labelsep=period}
\maketitle
\begin{sciabstract}
\vspace{-0.4cm}
In ordinary materials, electrons conduct both electricity and heat, where their charge-entropy relations observe the Mott formula and the Wiedemann-Franz law~\cite{mott1958theory,wiedemann1853relative}.
In topological quantum materials, the transverse motion of relativistic electrons can be strongly affected by the quantum field arising around the topological fermions~\cite{RevModPhys.82.1959}, where a simple model description of their charge-entropy relations remains elusive.
Here we report the topological charge-entropy scaling in the kagome Chern magnet TbMn$_6$Sn$_6$, featuring pristine Mn kagome lattices with strong out-of-plane magnetization.
Through both electric and thermoelectric transports, we observe quantum oscillations with a nontrivial Berry phase, a large Fermi velocity and two-dimensionality, supporting the existence of Dirac fermions in the magnetic kagome lattice.
This quantum magnet further exhibits large anomalous Hall, anomalous Nernst, and anomalous thermal Hall effects, all of which persist to above room temperature.
Remarkably, we show that the charge-entropy scaling relations of these anomalous transverse transports can be ubiquitously described by the Berry curvature field effects in a Chern-gapped Dirac model.
Our work points to a model kagome Chern magnet for the proof-of-principle elaboration of the topological charge-entropy scaling.

\end{sciabstract}


Elucidating the quantum nature and transport behaviors of topological magnets is at the frontier of condensed matter physics, and provides indispensable information on the application of quantum material research~\cite{keimer2017physics,sachdev2018topological}.
The interplay between topology and magnetism can naturally occur in kagome lattice systems~\cite{10.1143/ptp/6.3.306,Yin2021}.
Owing to its special lattice geometry, the electronic structure of kagome lattice inherently features both Dirac cones and band singularities including flat bands and van Hove singularities.
The former provides the source of topological nontrivial band structure, while the latter can drive magnetic instabilities.
In this regard, several transition metal based kagome magnets are of current interest, which exhibit many-body interplays and Chern quantum phases~\cite{Jiang2021,nakatsuji2015large,PhysRevLett.125.046401,liu2018giant,yin2019negative,Yin2020,
ye2018massive,yin2018giant,yin2020discovery,PhysRevLett.126.246602}.
Among these kagome magnets, TbMn$_6$Sn$_6$ stands out owing to its pristine kagome lattice (without other atoms in the kagome lattice plane) and strong out-of-plane magnetization (persisting to above room temperature).
Spatial and momentum resolved spectroscopies have uncovered the topological electronic structure that can be described by a simple Chern-gapped Dirac model~\cite{PhysRevLett.61.2015,PhysRevLett.95.226801,PhysRevLett.115.186802}$^,$\cite{yin2020discovery}.
The tight-binding model of kagome lattice considering the nearest neighbouring hopping hosts Dirac cones at the Brillouin zone corners.
Magnetic exchange interaction splits the spin-up and spin-down Dirac cones, while the combination of out-of-plane magnetization and Kane-Mele type spin-orbit coupling further opens the Chern gap for the spin-polarized Dirac fermions.
However, topological transport measurements of this kagome Chern magnet have been lacking.
Here we use the combination of electric, thermoelectric and thermal transport to characterize this quantum magnet and to discover its topological charge-entropy scaling.



TbMn$_6$Sn$_6$ crystallizes in the space group of P6/mmm, featuring a pure Mn-based kagome lattice.
Below its Curie temperature ($\sim$ 420~K), it possesses a ferrimagnetic state in which the magnetic moment of Tb is anti-parallel aligned with the ferromagnetic ordered Mn lattice due to the strong exchange coupling between Tb and Mn moments.
The ferrimagnetic state changes its anisotropy from an easy-plane to an out-of-plane configuration when the temperature is below 313~K.
The out-of-plane ferromagnetic Mn kagome lattice, which is stable over a wide range in the phase diagram (Fig.~1{a}), is crucial to support the fully spin-polarized Dirac fermions with a large Chern gap~\cite{yin2020discovery,PhysRevLett.126.246602}.
Fig.~1{b} shows the quantum oscillations resolved in the thermoelectric and electric transport measurements at low temperatures.
In the Seebeck signals, strong quantum oscillations with one dominant frequency ($\alpha$, 96~T) can be observed just above 5~T, which is the smallest field among all known kagome magnets.
The overall oscillatory patterns between thermopower and resistance show a $\pi/2$ phase shift, indicating that the $\alpha$ orbit is from a hole-like band~\cite{fletcher1981amplitude,matusiak2017thermoelectric}, consistent with spectroscopic observations~\cite{yin2020discovery}.
The cyclotron mass of $\alpha$ orbit is found to be $0.14\ m_e$ by fitting to the Lifshitz-Kosevich formula (See Method section).
According to the Onsager relation~\cite{shoenberg2009magnetic}, this frequency corresponds to a Fermi vector $k_F$ of 0.05~$\mathrm{\AA}^{-1}$.
The (Dirac cone) velocity $\nu_D=\hbar k_F/m^\ast$ is as large as $4.2\times10^5$~m/s, and the (Dirac cone) energy $E_D\simeq m^\ast\nu_D^2$ is estimated to be around 140~meV, in remarkable accordance with the Dirac dispersion observed in the previous spectroscopic experiments~\cite{yin2020discovery}.
We further subtract the Berry phase from oscillatory peak positions in Fig.~1{c}.
The intercept in the Landau fan diagram is found to be around -1/8, highlighting a nontrivial band topology of the $\alpha$ orbit~\cite{PhysRevLett.117.077201,murakawa2013detection}.

To confirm that the main frequency $\alpha$ is stemming from the quasi-two-dimensional electronic pocket from Mn kagome lattice, we performed magneto-Seebeck measurements when the magnetic field is tilted away from the $c$ axis.
As shown in Fig.~2{g}, both the oscillatory frequency $F$ and the cyclotron mass $m^\ast$ change in a $1/\cos\theta$ manner with respect to the tilted angle $\theta$.
For a gapped Dirac dispersion satisfying $E_D=\sqrt{(\hbar\nu_D k_F)^2+(\Delta/2)^2}$, the oscillatory frequency and cyclotron mass should follow the relations~\cite{Ye2019},
\begin{equation}
F=\frac{E_D^2-(\Delta/2)^2}{2e\hbar\nu_D^2\cos\theta},\ \ \  m^\ast=\frac{E_D}{\nu_D^2\cos\theta},
\end{equation}
where $\Delta$ is the gap size, $\nu_D$ is the Fermi velocity of the Dirac cone.
Because TbMn$_6$Sn$_6$ is a hard magnet with the easy axis pinned to the $c$ axis at low temperatures~\cite{VENTURINI199135,MALAMAN1999519,CLATTERBUCK199978,zhang2005unusual}, its $E_D$, $\Delta$ and $\nu_D$ change little when the external magnetic field is tilted away from the $c$ direction.
The ratio $F/m^\ast$ is always close to $E_D/2e\hbar$, strongly supporting that the main quantum oscillation orbit $\alpha$ in transport is stemming from the Chern-gapped Dirac fermion.
This is unlike the case of Fe$_3$Sn$_2$ whose electron mass and gap size can be easily tuned by an external magnetic field~\cite{yin2018giant,Ye2019,PhysRevLett.123.196604}.
The strong coupling between 4f and 3d electrons in TbMn$_6$Sn$_6$, which is absent in other transition-metal-bearing kagome magnets, guarantees a stable out-of-plane ferromagnetic Mn sublattice even in an extremely large external field and temperature range~\cite{zhang2005unusual}.
This large anisotropy makes the topological band of Chern-gapped Dirac fermion robustly against the change of magnetic field and the elevated temperature, providing an ideal platform for studying Chern related topology.

To further elucidate the quantum topology of this kagome magnet, we study the anomalous Hall effect, anomalous Nernst effect and anomalous thermal Hall effect.
These three effects, stemming from the electron's anomalous velocity endowed by the Berry curvature field in magnets~\cite{RevModPhys.82.1959}, are important fingerprints of the topological band structures residing near the Fermi energy.
Fig.~2{a-c} shows the Hall conductivity $\sigma_{xy}$, Nernst thermopower $S_{xy}$ and thermal Hall conductivity $\kappa_{xy}$, which behave consistently with the $M(H)$ loops at various temperatures shown in Fig.~S1{b}.
Below 220~K, sharp and large hysteresis loops can be observed in all three off-diagonal signals, corresponding to the spontaneous, out-of-plane ferrimagnetic state.
The hysteresis loops become narrow gradually and show additional bending as the in-plane magnetic components develop at higher temperature.
Above around 310~K, the three off-diagonal signals show the anomalous effects as long as the external field forces the in-plane moments to align along the $c$ axis~\cite{VENTURINI199135,MALAMAN1999519,CLATTERBUCK199978}.
The temperature dependence of the anomalous effects is summarized in Fig.~3{d-e}.
An anomalous Nernst coefficient as large as -2.4~$\mathrm{\mu V/K}$ is observed at 330~K, which is comparable to the largest values reported in other topological magnets~\cite{Sakai2018,PhysRevB.101.180404,yang2018giant,PhysRevX.9.041061}.
The anomalous thermal Hall conductivity on the other hand, reaches maximum value of 0.16~W/Km at 320~K.
These large anomalous transport terms at room temperature make TbMn$_6$Sn$_6$ a promising candidate for thermoelectric applications~\cite{doi:10.1080/14686996.2019.1585143,Sakai2020,https://doi.org/10.1002/adma.202100751}.

In order to further study the intertwined relationships between these topological transport behaviors, we analyse their scaling relations.
A conventional version in solids is the widely known Mott formula and the Wiedemann-Franz law, which associate the charge and entropy of the electron via~\cite{mott1958theory,wiedemann1853relative}
\begin{equation}
\bm{\alpha}=\frac{\pi^2}{3}\frac{k_B^2T}{e}\frac{\partial\bm{\sigma}}{\partial\varepsilon}|_{\varepsilon_F},
\end{equation}
\begin{equation}
\bm{\kappa}=\frac{\pi^2}{3}\frac{k_B^2}{e^2}\bm{\sigma} T=L_0\bm{\sigma}T,
\end{equation}
where $\varepsilon$ is the energy, $L_0=2.44\times10^{-8}V^2/K^2$ is the Sommerfeld value, $\bm{\sigma}$, $\bm{\alpha}$ and $\bm{\kappa}$ are the electric, thermoelectric and thermal conductivity tensor, respectively.
In the context of the anomalous transverse transport in topological magnets, however, their relevance remains much less explored.

As shown in Fig.~3{a}, we have confirmed that an intrinsic mechanism dominates in the anomalous Hall effect of TbMn$_6$Sn$_6$ by scaling $\sigma_{xy}^A$ versus $\sigma_{xx}$ in the presented temperature range~\cite{yin2020discovery,PhysRevLett.126.246602}.
A polynomial fitting~\cite{PhysRevLett.96.037204,PhysRevLett.103.087206} as shown in the inset of Fig.~3{a} gives an intrinsic contribution of $\sigma_{xy}^{int}\sim0.13\ e^2/h$ per kagome layer.
In accordance to the anomalous Hall conductivity and quantum oscillations of the $\alpha$ orbit, all parameters of the Chern-gapped Dirac fermion ($E_D\sim130~\mathrm{meV}$, $\Delta=2\sigma_{xy}^{int}E_Dh/e^2\sim34~\mathrm{meV}$) can been experimentally nailed down and its Berry spectrum can be depicted.
We can therefore trace the Berry curvature field to the Chern-gapped Dirac fermion in the stable out-of-plane ferrimagnetic state in TbMn$_6$Sn$_6$, which gives a unique opportunity to test the scaling relation of the $\sigma_{xy}^A$, $\alpha_{xy}^A$ and $\kappa_{xy}^A$ of the Chern-gapped Dirac fermion in a topological magnet.

The anomalous thermoelectric Hall conductivity is calculated as $\alpha_{xy}^A=\sigma_{xy}^AS_{xx}+\sigma_{xx}S_{xy}^A$.
As shown in Fig.~3{c}, the ratio $\alpha_{xy}^A/\sigma_{xy}^A$ in TbMn$_6$Sn$_6$ is small at low temperatures, and then shows an overall increase at elevated temperatures, reaching $-67~\mathrm{\mu V/K}$ at 330~K.
For a Chern-gapped Dirac model, it can be treated analytically that $\partial\sigma_{xy}^A/\partial\varepsilon=\sigma_{xy}^A/E_D$.
Substituting this into the Mott formula in Eq.~(2), we have the relation between $\alpha_{xy}^A$ and $\sigma_{xy}^A$,
\begin{equation}
\alpha_{xy}^A/\sigma_{xy}^A=\frac{\pi^2}{3}\frac{k_B}{e}\frac{k_BT}{E_D}.
\end{equation}
The experimental data fit well into this simple linear line.
The deviation around 100~K is currently unclear, and we speculate its origin from spin excitations~\cite{doi:10.1126/science.1257340,PhysRevLett.115.106603} as also evident in the magnetization susceptibility in Fig.~S1 which deserves future study.
It has been suggested empirically that the ratio $\alpha_{xy}^A/\sigma_{xy}^A$ should be a sizable fraction of $\left|k_B/e\right|\sim 86~\mathrm{\mu V/K}$ in topological magnets~\cite{PhysRevX.9.041061,PhysRevB.101.180404}.
Our analysis shows that for the kagome Chern magnet, the physics behind the $k_B/e$ threshold is actually a competition between two energies: the thermal energy $k_BT$ and the Dirac cone energy $E_D$.

We now study the relation between the anomalous thermal Hall conductivity $\kappa_{xy}^A$ and anomalous Hall conductivity $\sigma_{xy}^A$.
As shown in Fig.~3{d}, the ratio $\kappa_{xy}^A/\sigma_{xy}^A$ apparently deviates from the linear $T$ dependence expected by the standard Wiedemann-Franz law.
It apparently surpasses $L_0T$ above 100~K, reaching 1.8 times $L_0T$ at 320~K and then the slope weakly damps at higher temperature.
This substantial deviation can not be attributed to experimental errors, as the heat loss and geometric factors would only introduce at most 10\% uncertainty in our study.
It was reported that $\kappa_{xy}^A/\sigma_{xy}^AT$ remains close to $L_0$ in Mn$_3$Sn and Co$_2$MnGa when the Berry spectrum is smooth in the vicinity of the Fermi level~\cite{PhysRevLett.119.056601,PhysRevB.101.180404}.
In Ni and Fe, the $\kappa_{xy}^A/\sigma_{xy}^AT$ deviates the Sommerfeld value downwards at higher temperatures due to the inelastic scattering ~\cite{PhysRevLett.100.016601,doi:10.1126/sciadv.aaz3522}.
A particular Berry spectrum also causes a suppressed $\kappa_{xy}^A/\sigma_{xy}^AT$ at high temperatures in Mn$_3$Ge and Co$_3$Sn$_2$S$_2$~\cite{doi:10.1126/sciadv.aaz3522,Ding_2021}.
To the best of our knowledge, TbMn$_6$Sn$_6$ is the first topological magnet that possesses an enhanced $\kappa_{xy}^A/\sigma_{xy}^AT$ ratio at elevated temperatures.
The inelastic scattering mechanism is self-evidently ruled out as it should cause an inverse effect.
Considering the phonon Hall and magnon Hall effect is usually in the order of 10$^{-3}$~W/Km~\cite{PhysRevLett.95.155901,doi:10.1126/science.1188260,PhysRevLett.115.106603,doi:10.1126/science.1257340}, they are too small to account for this substantial enhancement.
Below we show this anomaly in TbMn$_6$Sn$_6$ is due to the Berry curvature field effect hosted in the Chern-gapped Dirac fermions.

In transport theories, $\sigma_{xy}^A$ and $\kappa_{xy}^A$ can be expressed in the form of the Berry curvature $\Omega_{z}^n(\bm k)$~\cite{PhysRevB.64.224508},
\begin{equation}
\sigma_{xy}^A=\frac{e^2}{\hbar}\int d\varepsilon  \left(-\frac{\partial f}{\partial \varepsilon }\right)\tilde\sigma_{xy}({\varepsilon }),
\end{equation}
\begin{equation}
\kappa_{xy}^A=\frac{1}{\hbar T}\int d\varepsilon  \left(-\varepsilon ^2\frac{\partial f}{\partial \varepsilon }\right)\tilde\sigma_{xy}({\varepsilon}),
\end{equation}
\begin{equation}
\tilde\sigma_{xy}({\varepsilon})=\int_{BZ}\frac{d\bm k}{(2\pi)^3}\sum_{\varepsilon_n<\varepsilon }\Omega_{z}^n(\bm k),
\end{equation}
where $f$ is the Fermi-Dirac distribution.
In a Chern-gapped Dirac model, $\tilde\sigma({\varepsilon })$ can be calculated as~\cite{PhysRevB.75.045315}
\begin{equation}
\tilde\sigma_{xy}({\varepsilon})=\frac{1}{2\pi}\frac{\Delta}{\sqrt{\Delta^2+4\hbar^2 k_F^2 v_D^2}}=\left\{
\begin{aligned}
&\frac{1}{2\pi}\frac{\Delta/2}{\left|\varepsilon-E_D\right|}&, \quad \left|\varepsilon-E_D\right|>\Delta/2\\
&\qquad \frac{1}{2\pi}&, \quad \left|\varepsilon-E_D\right|<\Delta/2.
\end{aligned}
\right.
\end{equation}
As shown in Fig.~3{b}, the pondering functions in $\sigma_{xy}^A$ and $\kappa_{xy}^A$, namely $-\partial f/\partial \varepsilon $ and $-\varepsilon ^2\partial f/\partial \varepsilon $ have different shapes.
The former is a delta-like function, while the latter adopts a double-peak-like profile and is more extended in the energy scale.
The Wiedemann-Franz law would only hold when $\tilde\sigma_{xy}({\varepsilon})$ is antisymmetric around zero energy.
However in the case of a Chern-gapped Dirac fermion, Berry curvature maximizes around the Dirac gap, and contributes more in the latter integration as $-\varepsilon ^2\partial f/\partial \varepsilon $ lies more close to the gap with increasing temperature.
We note that it is the singularity of Berry curvature from the Chern Dirac gap that results in the enhancement of $\kappa_{xy}^A/\sigma_{xy}^A$.
It can be estimated that the most deviation from $T$-linearity occurs around $E_D/5k_B$, namely 300~K in our case.
The experimental ratios of $\kappa_{xy}^A/\sigma_{xy}^A$ fit well into the calculated curves from the simple model.

Due to the complex of magnetism and thermoelectric/thermal transport in real magnetic materials, it has been difficult to relate the anomalous Nernst effect and anomalous thermal Hall effect directly with the topological electrons.
A common strategy is to calculate the energy dependence of the Berry curvature from ab initio and thoroughly scan over the energy to find a rough match with the anomalous transverse effects in experiment.
With the lack of the topological electrons' detailed information, the validity of this strategy, to a large extent, depends on the accuracy of the ab initio calculation which is known to be difficult for correlated magnetic materials.
In this work we have, for the first time, established a direct link between the anomalous effects and a particular topological fermion in a kagome magnet TbMn$_6$Sn$_6$.
Starting from the angle-dependent quantum oscillations, we depict the two-dimensional Dirac dispersion.
Then by comparing the anomalous transverse electric, thermoelectric and thermal transports, we are able to connect the anomalous effects to the Berry spectrum of the Chern-gapped Dirac fermion.
In particular, two ratios, the $\alpha_{xy}^A/\sigma_{xy}^A$ and $\kappa_{xy}^A/\sigma_{xy}^A$, give inherent scaling relations.
The former scales with $k_BT/E_D$ over an extended temperature range, while the latter enhances over the linear $T$-dependence above 100~K.
Only two parameters, the Dirac cone energy $E_D$ and the gap size $\Delta$ both of which are determined from experiments, are involved in our discussion, and the simple Chern-gapped Dirac model is able to capture most of the (anomalous) transport behaviors.
Our results, therefore, establish a new model topological magnet for the proof-of-principle elaboration of the topological charge-entropy scaling.
In future, it would be interesting to engineer this material down to atomic layers to realize higher temperature quantum anomalous Hall effect and to explore the topological scaling at the quantum-limit.


\clearpage

\begin{figure}[htbp]
\begin{center}
\includegraphics[clip, width=1\textwidth]{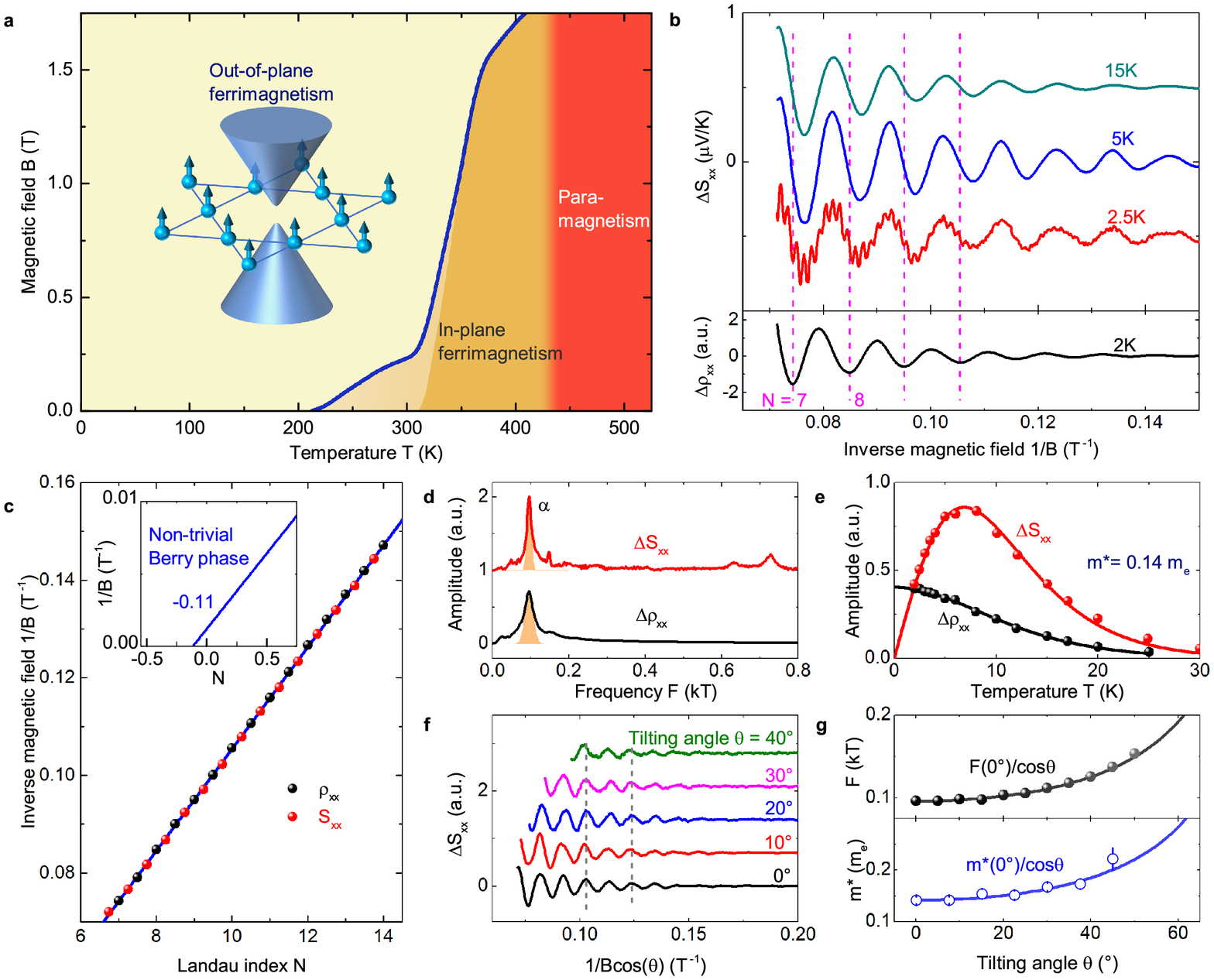}\\[0pt]  
\caption{{\bf Topological quantum oscillations.}
{\bf a,} Magnetic phase diagram of TbMn$_6$Sn$_6$ when the field $\bm B$ is applied along the crystallographic $c$ axis.
Three main regions can be resolved, including the out-plane ferrimagnetic state where Chern-gapped Dirac states are supported, in-plane ferrimagnetic state, and paramagnetic state above 420~K.
Inset shows the magnetic Mn kagome lattice and corresponding Chern-gapped Dirac cone at the Brillouin Zone corner of the momentum space.
{\bf b,} Quantum oscillations revealed in magneto-Seebeck signal $S_{xx}$ and magnetoresistance $\rho_{xx}$.
There is a $\pi/2$ phase shift between oscillatory parts of $S_{xx}$ and $\rho_{xx}$.
{\bf c,} Landau fan diagram for the oscillations, suggesting a non-trivial Berry phase.
{\bf d,} Fast Fourier transform spectrum of the oscillatory component in $S_{xx}$ and $\rho_{xx}$, showing dominant contribution of the $\alpha$ orbit.
{\bf e,} Cyclotron mass fitting of oscillatory amplitude to the Lifshitz-Kosevich formulas.
{\bf f,} Oscillatory parts in Seebeck signals when the field is tilted away from the $c$ direction.
{\bf g,} Angle dependent oscillatory frequencies and corresponding cyclotron mass for orbit $\alpha$, both showing an inverse cosine behavior.
}
\label{f1}
\end{center}
\end{figure}
\clearpage

\begin{figure}[htbp]
\begin{center}
\includegraphics[clip, width=0.95\textwidth]{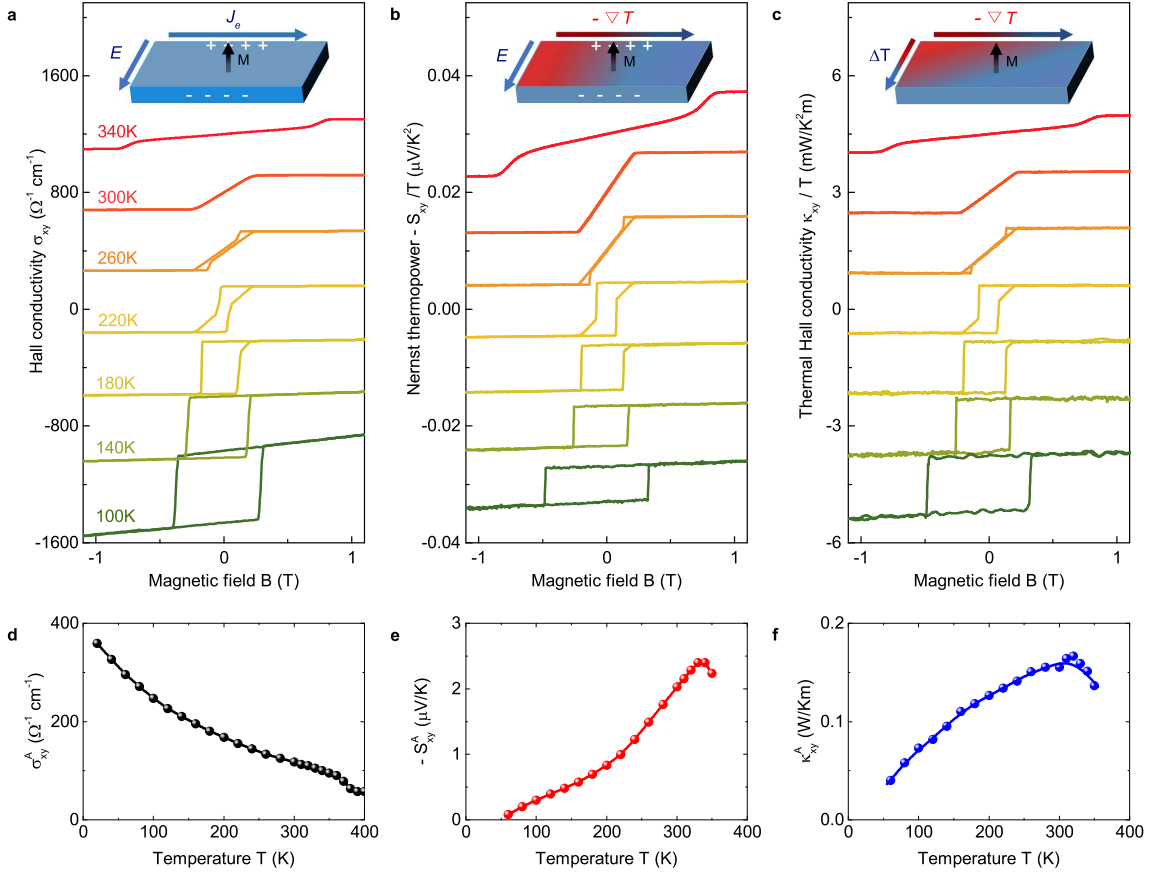}\\[0pt]  
\caption{{\bf Topological transverse transport.}
{\bf a-c,} The electric Hall conductivity $\sigma_{xy}$, the Nernst thermopower divided by temperature $-S_{xy}/T$ and thermal Hall conductivity divided by temperature $\kappa_{xy}/T$ at representative temperatures, showing dominant contribution of anomalous terms. Curves are shifted vertically for clarity.
The resemblance of $\sigma_{xy}$, $S_{xy}$ and $\kappa_{xy}$ profiles indicates a shared origin from Berry curvature contributions.
Insets show the sketches for anomalous Hall, Nernst and thermal Hall effect.
{\bf d-f,} The temperature dependence of $\sigma_{xy}^A$, $S_{xy}^A$ and $\kappa_{xy}^A$, respectively.
}
\label{f2}
\end{center}
\end{figure}

\begin{figure}[htbp]
\begin{center}
\includegraphics[clip, width=0.9\textwidth]{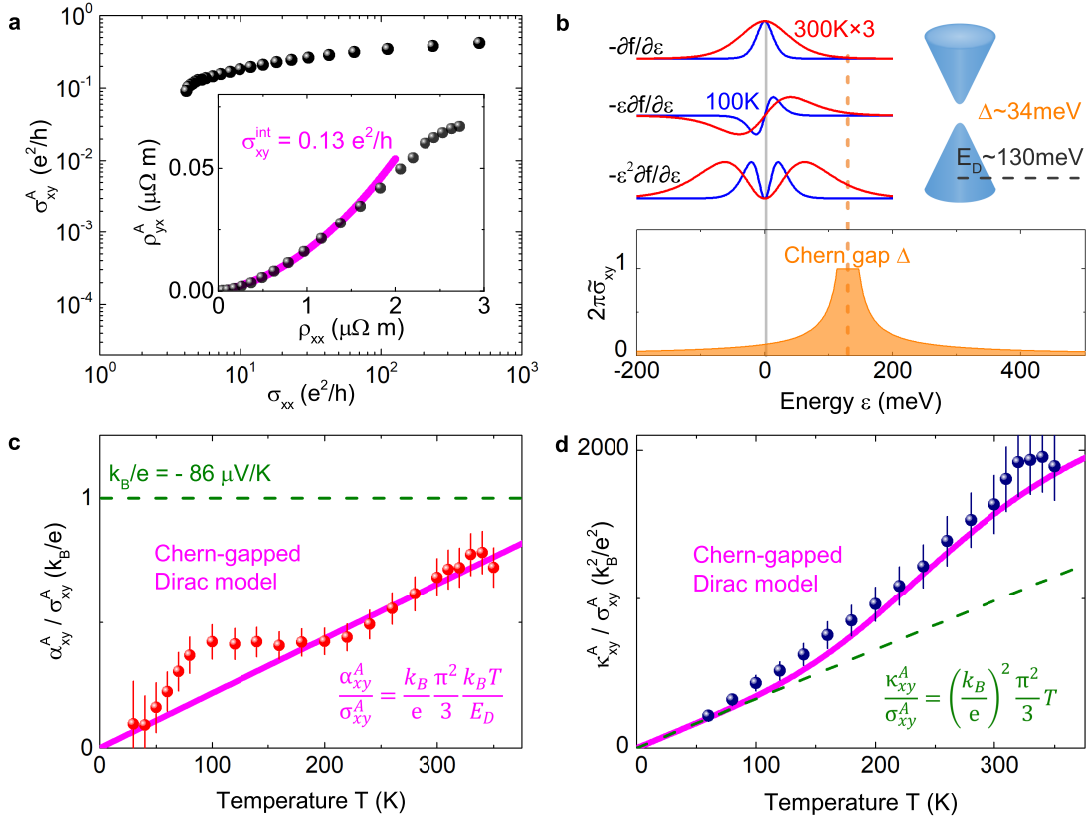}\\[0pt]  
\caption{{\bf Topological charge-entropy scaling.}
{\bf a,} Scaling of the anomalous Hall conductivity. The longitudinal conductivity $\sigma_{xx}$ for TbMn$_6$Sn$_6$ lies within the good metal region, suggesting a dominant intrinsic contribution. Inset shows a polynomial fitting of the intrinsic Hall conductivity, amounting to 0.13~$e^2/h$ per kagome layer.
{\bf b,} The pondering function for anomalous Hall conductivity $\sigma_{xy}^A$, anomalous thermoelectric Hall conductivity $\alpha_{xy}^A$, and the anomalous thermal Hall conductivity $\kappa_{xy}^A$ at 100~K and 300~K, respectively, together with the $2\pi\tilde\sigma_{xy}({\varepsilon})$ for the Chern gapped Dirac fermion with a gap size of $\Delta$. For clarity, the pondering functions at 300~K are multiplied by 3 times.
Top-right sketch shows the Chern-gapped Dirac cone with gap size $\Delta\sim$ 34~meV and Dirac cone energy 130~meV.
{\bf c,} The ratio $\alpha_{xy}^A/\sigma_{xy}^A$ at different temperatures, approaching $k_B/e$ at around 330~K.
This ratio scales with the linear function of $k_BT/E_D$, which is obtained from the Chern-gapped Dirac model.
{\bf d,} The ratio $\kappa_{xy}^A/\sigma_{xy}^A$ at different temperatures. Above 100~K, this ratio significantly enhances over the $T$-linear function expected by the Wiedemann-Franz law at elevated temperatures, which matches the behavior for the Chern-gapped Dirac model.
}
\label{f3}
\end{center}
\end{figure}

\clearpage

\end{document}